\documentclass[12pt]{article}

\usepackage{latexsym}
\usepackage{amssymb}
\usepackage{amsmath}
\usepackage{amsfonts}
\usepackage{mathrsfs}
\usepackage{cite}
\usepackage{axodraw}
\usepackage{graphics}
\usepackage{graphicx,array}
\numberwithin{equation}{section}

\def\m{{\mu}}

\def\a{{\alpha}}

\def\b{{\beta}}
\def\g{{\gamma}}

\def\d{{\delta}}

\def\t{{\theta}}

\def\ep{{\varepsilon}}

\def\f{\frac}

\def\la{\label}
\def\eq{\eqref}
\def\pr{\partial}

\newcommand{\be}{\begin{equation}}
\newcommand{\ee}{\end{equation}}

\newcommand{\Tr}{{\rm Tr}}

\def\cN{{\cal N}}
\def\bea{\begin{eqnarray}}
\def\eea{\end{eqnarray}}
\def\nn{\nonumber}

\begin{document}

\begin{titlepage}
\phantom{x}

\vspace{1cm}

\begin{center}

{\Large \bf On effective K\"ahler potential in
$\cN=2$, $d=3$ SQED  \\[3mm]
} \vspace{1.5cm}
 {\bf
 I.L. Buchbinder $^+$\footnote{joseph@tspu.edu.ru},
 B.S. Merzlikin $^{+\dag}$\footnote{merzlikin@tspu.edu.ru},
\\[3mm]
 {\it $^+$ Department of Theoretical Physics, Tomsk State Pedagogical
 University,\\ Tomsk 634061, Russia\\
 and\\ National Research Tomsk State University, Tomsk, 634050 Russia
\vskip 0.08cm \vskip 0.08cm
 $^\dag$
 Department of Higher Mathematics and Mathematical Physics,\\
\it Tomsk Polytechnic University, 634050 Tomsk, Russia}
 \\[1cm]
}

\begin{abstract}
We compute the two-loop effective K\"ahler potential in
three-dimensional $\cN=2$ supersymmetric electrodynamics with
Chern-Simons kinetic term for the gauge superfield. The effective
action is constructed on the base of background field method with
one parametric family of gauges. In such an approach, the quadratic
part of quantum action mixes the gauge and matter quantum
superfields yielding the complications in the computations of the
loop supergraphs. To avoid this obstacle and preserve dependence on
the gauge parameter we make a nonlocal change of quantum matter
superfields after which the propagator is diagonalized, however the
new vertices have appeared. We fix the suitable background and
develop the efficient procedure of calculating the two-loop
supergraphs with the new vertices. We compute the divergent and
finite parts of the superfield effective action, find the two-loop
effective K\"ahler potential and show that it does not depend on the
gauge parameter.

\end{abstract}
\end{center}

\end{titlepage}
\setcounter{footnote}{0}

\section{Introduction}

It is well known that the leading part of the low-energy effective
action in the supersymmetric field models with chiral matter
superfields is described by effective K\"ahler potential (see, e.g.,
\cite{book}). The effective K\"ahler potential is responsible for
the structure of the quantum moduli space of $4D$, $\cN=1$
gauge-matter theories in the Higgs branch and is closely related to
supersymmetric sigma-models. Computations of the effective K\"ahler
potential in the $4D$, $\cN=1$ supersymmetric models have been
carried out in many papers (see
e.g.\cite{BKY,BKPY,BKP,BCP1,BCP2,Brignole},
\cite{deWit1996,PW,Grisaru96} for one-loop calculations and
\cite{Nibbelink} for two-loop calculations)\footnote{The detailed
analysis of the $4D$ superfield effective potentials has been given
in the thesis \cite{Tyler}.}.

In three-dimensional supersymmetric theories the structure of
effective K\"ahler potential is much less well understood. The
effective superpotential in $\cN=1$ gauge-matter theories was
studied in \cite{Petrov,Lehum}, but it does not correspond to
K\"ahler sigma-models for component scalar fields due to an
insufficient number of supersymmetries. The two-loop effective
K\"ahler potential was computed for the three-dimensional
Wess-Zumino model in $\cN=2$ superspace \cite{BMS3} but it has not
been studied in gauge-matter models which have much more interesting
classical and quantum properties. Note that there is a broad
discussion of the structure of moduli space of three-dimensional
gauge theories with $\cN=2$ supersymmetry including its Higgs branch
(see, e.g., \cite{deBoer97,deBoer,AHISS,IS13}), but the
corresponding K\"ahler potential has never been computed explicitly
in perturbation theory. On the contrary, the low-energy effective
action in the Coulomb branch of three-dimensional gauge-matter
models has recently been studied in the $\cN=2$ superspace up to
two-loop order \cite{1,2,3,4,5}.

The aim of the present paper is to initiate the study of the
perturbative quantum corrections to the effective K\"ahler potential
in three-dimensional $\cN=2$ supersymmetric gauge theories. We
compute two-loop effective K\"ahler potential in three-dimensional
$\cN=2$ supersymmetric quantum electrodynamics (SQED) with
Chern-Simons kinetic term for the gauge superfield. At the classical
level this model is superconformal, but we show that the conformal
invariance is broken by two-loop quantum corrections. We find that
the two-loop K\"ahler potential in the $\cN=2$ supersymmetric
electrodynamics is similar in some aspects to the one-loop effective
K\"ahler potential in four-dimensional $\cN=1$ SQED \cite{PW}.

In the present paper we study the effective K\"ahler potential for
one particular model: $\cN=2$ SQED with the Chern-Simons kinetic
term for the gauge superfield and two chiral superfields. The
following are arguments supporting the study the effective K\"ahler
potential in this model:
\begin{itemize}
\item Although this model is quite simple, it possesses a
non-trivial effective K\"ahler potential which represents the
leading part of the low-energy effective action in the Higgs branch.
Note that, in contrast to the four-dimensional case, we need to
study the two-loop effective action since, as we will show further,
the one-loop quantum corrections to the effective K\"ahler potential
are trivial in the sense that they repeat the form of classical
K\"ahler potential. In general, computation of two-loop quantum
corrections is a hard routine, but in the present case we need to
consider just a few two-loop Feynman graphs since the model is
Abelian and, in particular, ghost superfields do not contribute.
\item As we will show further, the form of two-loop quantum
corrections to the effective K\"ahler potential is in fact dictated
by logarithmically divergent supergraphs. Hence, the effective
K\"ahler potential which is proper to $\cN=2$ Chern-Simons-matter
theories it seems can not appear in three-dimensional models such as
$\cN>2$ Chern-Simons-matter gauge theories which have no UV
divergences \cite{Avdeev1,Avdeev2,BILPSZ} or the $\cN=2$ SQED with
Maxwell kinetic term for the gauge superfield which is
superrenormalizable.
\item The $\cN=2$ SQED with the Chern-Simons kinetic term is
classically superconformal, but, as we will show, the two-loop
quantum corrections to the low-energy effective action break the
conformal invariance. This is analogous to the holomorphic
low-energy effective action in four-dimensional $\cN=2$ gauge
theories \cite{Seib88} which is known to be responsible for the
superconformal symmetry breaking.
\item We consider the $\cN=2$ SQED with two chiral
superfields having different charges with respect to the gauge
superfield. This model is advantageous as compared to similar models
with odd number of chiral superfields which may have parity anomaly
\cite{anomaly1,anomaly2,anomaly3}. Moreover, in the considered model
the effective K\"ahler potential can be unambiguously computed
within the background field method since we can fix the background
for chiral matter superfields which solves classical equations of
motion. As a result, the obtained effective K\"ahler potential
corresponds to the gauge-independent part of the effective action.
\end{itemize}

Let us discuss several technical points concerning two-loop
computations in the considered model. The effective action in
quantum field theory of gauge fields is known to be a
gauge-dependent quantity. However, the effective action calculated
for background field satisfying the effective equations of motions
is gauge independent (see e.g. \cite{DeWitt}). When we study the
perturbative quantum corrections to effective action in the frame of
loop expansion, the gauge independent one-loop corrections should be
considered on the classical equations of motion while for the gauge
independent two-loop quantum corrections we have to take into
account the effective equations of motion up to one-loop order. In
the $\cN=2$ SQED studied in the present paper it is sufficient to
consider constant background chiral superfields to compute the
effective K\"ahler potential. As we will demonstrate, this
background obeys not only classical but also quantum effective
equations of motion up to one-loop order. This guarantees that the
two-loop effective K\"ahler potential computed in this model is
gauge independent. Moreover, in the functional integral we fix the
gauge freedom, but we keep the gauge-fixing parameter $\alpha$
arbitrary throughout all quantum computations. We directly
demonstrate that the obtained one- and two-loop quantum corrections
to the effective K\"ahler potential are independent of $\alpha$,
confirming its gauge independence.

Another technical comment concerns the details of applications of
the background field method at the two-loop order. When we perform
the background quantum splitting the classical action acquires a
number of terms which mix gauge and matter superfields at the
quadratic order and make the propagator non-diagonal. In quantum
computations it is desirable to deal with the diagonal propagator
for quantum superfields. Otherwise the computations become extremely
complicated. There are, in general, two ways to achieve this: (i) to
make a non-local change of fields to diagonalize the propagator or
(ii) to apply a generalized gauge-fixing condition ($R_\xi$-gauge)
which eliminates the mixed terms at the quadratic order. The latter
approach is usually simpler, but it does not allow one to keep the
gauge-fixing parameter arbitrary. Therefore, in the present work we
make a non-local change of quantum superfields to bring the
propagator to the diagonal form. The cost for this is that we get
new interaction vertices having non-local form and playing important
role in two-loop quantum computations. This means we should develop
a specific technique to compute the supergraphs with the new
vertices.

The rest of this paper is organized as follows. Section 2 is devoted
to some preliminary discussion concerning the structure of loop
quantum corrections to the effective K\"ahler potential and specify
the background which is suitable for its evaluation. In section 3 we
perform the background-quantum splitting and derive the form of
propagators and interaction vertices which will be employed in loop
quantum computations. In the next two sections we calculate one- and
two-loop quantum effective actions and derive the form of effective
K\"ahler potential at the two-loop order. In the last section we
discuss the possible extensions of the results of the present paper.
In the appendices we collect some technical details of two-loop
quantum computations. Throughout this paper we use the $\cN=2$,
$d=3$ superspace notations and conventions introduced in earlier
works \cite{1,2,3,4,5}.


\section{Classical action and specification of background}

We consider the three-dimensional $\cN=2$ supersymmetric
electrodynamics which is described by two chiral matter superfields
$Q_+$ and $Q_-$ and a gauge superfield $V$ with superfield strength
$G=\frac i2 \bar D^\alpha D_\alpha V$. In general at a classical
level such a  model can have several parameters: the complex and
real mass parameters of the chiral superfields, the topological mass
of the gauge superfield and the Fayet-Iliopoulos term. In the
present paper we consider a particular case where the masses of
chiral superfields are vanishing and the gauge superfield has
infinite topological mass. Moreover, when the Fayet-Iliopoulos term
vanishes, the model is superconformal at the classical level. The
only parameter in the classical action is the Chern-Simons level $k$
\be
  S=\frac{k}{2\pi}\int d^7 z\, VG - \int d^7
z\,(\bar Q_+ e^{2V} Q_+ + \bar Q_- e^{-2V} Q_-)\,. \label{2CS0} \ee
The corresponding classical equations of motion are
\begin{subequations}
\label{eom} \bea
&&G=\frac{2\pi}k (\bar Q_+ e^{2V}Q_+ - \bar Q_- e^{-2V} Q_-)\,,\\
&&\bar D^2(\bar Q_\pm e^{\pm2V}) =0\,,\qquad
 D^2(Q_\pm e^{\pm 2V})=0\,.
\eea
\end{subequations}

The most natural approach to study the quantum effective action is
the background field method. For gauge theories in the $\cN=1$ $d=4$
superspace this method is discussed in \cite{GGRS}. Basic features
of this method for $\cN=2$ $d=3$ superspace were formulated in
\cite{1,2,3}. For recent applications of this method for computing
the effective actions in three-dimensional gauge theories in the
sector of gauge superfield see \cite{4,5}. Following this method, we
split the original superfields $V$, $Q_\pm$ and $\bar Q_\pm$ into
the so-called `quantum' $v$, $q_\pm$, $\bar q_\pm$ and
`background' $V$, $Q_\pm$, $\bar Q_\pm$ parts%
\footnote{Note that we denote the background superfields by the same
letters as the original ones. We hope that this will not lead to
misunderstandings since after the background-quantum splitting the
original superfields never appear.} \be V \to v + V\,,\quad Q_{\pm}
\to q_\pm + Q_{\pm} \,, \quad \bar Q_\pm \to \bar q_{\pm} + \bar
Q_{\pm}\,. \label{2split1} \ee Then, after fixing the gauge freedom
for the quantum gauge superfield $v$, one obtains a gauge-invariant
effective action for the background superfields \be
\Gamma=\Gamma[V,Q_\pm,\bar Q_\pm]\,. \ee

In the resent paper, we restrict ourselves to considering only the
part of the effective action which is described by the effective
K\"ahler potential for the chiral superfields $K_{\rm
eff}(Q_\pm,\bar Q_\pm)$. For this purpose it is sufficient to
consider the background superfields subject to the following
constraints \be V=0\,,\quad D_\alpha Q_\pm =0\,,\quad \bar D_\alpha
\bar Q_\pm =0\,. \label{bg1} \ee

In general, the effective action is gauge dependent but, if the
background superfields satisfy the exact effective equations of
motions, the effective action is gauge independent (see e.g.
\cite{DeWitt}). To get the gauge independent  one-loop effective
action it is sufficient to use the background superfields obeying
the classical equations of motion. For the two-loop effective action
we should take the background fields satisfying the one-loop
equations of motion. In the case under consideration we assume that
the background superfields obey not only (\ref{bg1}), but also
(\ref{eom}), \be \bar Q_+ Q_+ = \bar Q_- Q_- \equiv \bar Q Q\,.
\label{bg2} \ee In principle, this constraint could be modified by
one-loop corrections, but as we will show further, this is not the
case and it can be safely used for two-loop computations as well.
Thus, the problem is reduced to finding the effective action which
is described by a single function $K_{\rm eff}(\bar Q Q)$ \be
 \Gamma=-\int d^7z\, K_{\rm eff}(\bar Q Q)\,.
\ee We emphasize that this part of the effective action is gauge
independent and can be unambiguously computed in the two-loop
approximation for the background superfields constrained by
(\ref{bg1}) and (\ref{bg2}).

In the present paper we will compute the effective K\"ahler
potential up to two-loop order in the quantum perturbation theory
\be K_{\rm eff}=K+K^{(1)}+K^{(2)}+\ldots \label{K} \ee Here $K$ is
the classical K\"ahler potential, $K = 2 \bar Q Q$ while $K^{(1)}$
and $K^{(2)}$ correspond to one- and two-loop quantum contributions.
The ellipsis stand for higher loop quantum corrections which are
beyond our considerations.

\section{Propagators and vertices}
\label{Sect-props}

Upon the background-quantum splitting (\ref{2split1}), the classical
action (\ref{2CS0}) is  expanded in a series over the quantum
superfields, \be S= S_0 +  S_1 + S_2 + S_{\rm int}+ \ldots\,,
\label{decomposition} \ee where $S_0$ is the action depending only
on the background superfields and having the form (\ref{2CS0});
$S_1$ is liner in quantum superfields part and does not give rise to
one-particle irreducible diagrams for effective action. The action
$S_2$ is quadratic in quantum superfields \bea S_{2}&=&\int d^7 z\,
v\left( \f{ik}{4\pi}\bar D^\a D_\a + M\right)v -  \int d^7
   z\,(\bar q_+ q_+ + \bar q_- q_-) \nn \\
   && -\, 2\int d^7 z\,(\bar Q_+ \, q_+ v + Q_+\, \bar q_+ v - \bar Q_- \,
   q_- v - Q_-\, \bar q_- v)\,,
\label{SCS2} \eea while $S_{\rm int}$ is responsible for cubic and
quartic vertices for quantum superfields, \bea S_{\rm int}&=& - \int
d^7 z \Big(2\bar Q_+\, q_+ v^2 +
   2 Q_+\, \bar q_+ v^2 + 2\bar Q_-\, q_- v^2 + 2 Q_-\, \bar q_- v^2  \nn \\
   && +\,2 \bar q_+ q_+ v - 2 \bar q_-  q_- v
   + 2\bar q_+ q_+ v^2 + 2 \bar q_-  q_- v^2  -\f13 M v^4  \nn\\
   && + \f43 \bar q_+ v^3 + \f43 q_+ v^3 - \f43 \bar q_- v^3  -\f43 \bar q_- v^3
   -\f43 (\bar Q_+ Q_+ - \bar Q_- Q_-) v^3     \Big)\,,
\label{Sint} \eea where \be M \equiv -2(\bar Q_+ Q_+ + \bar Q_- Q_-)
= -4\, \bar Q Q\,. \label{M} \ee The ellipses in
(\ref{decomposition}) stand for higher order interaction vertices
for quantum superfields which are irrelevant for two-loop
computations.

The operator $\bar D^\alpha D_\alpha$ in (\ref{SCS2}) is degenerate
and requires gauge fixing. We use standard gauge fixing conditions
in the $\cN=2$ $d=3$ superspace \be \bar D^2 v =0\,, \qquad  D^2
v=0\,, \label{gauge-fixing} \ee which can be effectively taken into
account by adding to (\ref{SCS2}) the following gauge-fixing action
\cite{Avdeev1,Avdeev2,NG} \be
 S_{\rm gf} = \frac{ik\alpha}{8\pi}\int
d^7z\, v(D^2 + \bar D^2) v\,, \label{2Sgf} \ee where $\alpha$ is a
real parameter. In the present paper we do not fix this parameter
and keep it arbitrary. As we will show further, the effective
K\"ahler potential is independent of this parameter. It means in
fact that we study the gauge-independent part of the effective
action.

We consider the Abelian gauge theory, therefore the ghost
superfields, associated with the gauge fixing (\ref{gauge-fixing}),
decouple and do not contribute to the effective action.

Note that the quadratic action $S_2$ contains the mixed gauge and
matter quantum field terms given in the second line of (\ref{SCS2}).
This unpleasant feature leads to non-diagonal propagator for the
quantum superfields and makes quantum loop computations more
involved. However, it is always possible to make a non-local change
of (anti)chiral superfields in the functional integral such that the
propagator becomes diagonal. For the action (\ref{SCS2}) such a
change of fields reads \footnote{A similar non-local change of
fields was used in \cite{Kuz07} within computations of one-loop
effective K\"ahler potential in four-dimensional $\cN=1$ SQED (see
also earlier paper \cite{OV} for non-local change of fields in
non-supersymmetric QED).}
 \bea
 q_{+}(z)&\to& q_+(z) + 2\int d^7 z' G_{+-}(z,z') Q_+(z') v(z') \,, \nn\\
 q_{-}(z)&\to& q_-(z) - 2\int d^7 z' G_{+-}(z,z') Q_-(z') v(z') \,, \nn\\
\bar q_{+}(z)&\to& \bar q_{+}(z) + 2\int d^7 z' G_{-+}(z,z')\bar
Q_{+}(z') v(z')
\,,\nn\\
\bar q_{-}(z)&\to& \bar q_{-}(z) - 2\int d^7 z' G_{-+}(z,z')\bar
Q_{-}(z') v(z')\,,
 \label{2chov2}
\eea where $G_{+-}$ and $G_{-+}$ are propagators for the
(anti)chiral superfields obeying the equations
 \be
 \frac14 D^2 G_{+-}(z,z')=-\delta_-(z,z')\,,\qquad
 \frac14 \bar D^2 G_{-+}(z,z')=-\delta_+(z,z')\,.
\label{chir-props}
 \ee
Here $\delta_\pm(z,z')$ are chiral and anti-chiral delta-functions
which are related to the full superspace delta-function
$\delta^7(z-z')$ as
 \be
 \delta_+(z,z') = -\frac14\bar D^2
 \delta^7(z-z')\,,\qquad \delta_-(z,z') = -\frac14 D^2
 \delta^7(z-z')\,. \ee
The propagators $G_{+-}$ and $G_{-+}$ obeying (\ref{chir-props})
have the following explicit form
 \be
 G_{+-}(z,z')=\f{1}{\square} \f{\bar D^2
 D^2}{16}\d^7(z-z')\,,\qquad
 G_{-+}(z,z')=\f{1}{\square} \f{ D^2 \bar D^2}{16}\d^7(z-z')\,.
 \label{G+-def}
 \ee
Indeed, after the change of fields (\ref{2chov2}) the action
(\ref{SCS2}) acquires the form
 \bea
 S_2+S_{\rm gf} &=&  \int d^7z\, v \left(\f{ik}{4\pi}\bar D^\a D_\a +\f{ik\a}{8\pi}(\bar D^2 + D^2) +M\right) v -  \int d^7
z\,(\bar q_+ q_+ + \bar q_- q_-)
\nn\\
 && -\int d^7z\, d^7z'\, v(z) v(z') M
  \f{\{\bar D^2, D^2\}}{16\square}\d^7(z-z')\,.
\label{2SCS3}
 \eea
Here, in the last term, we used the fact that the spinor derivatives
do not act on the background superfields according to (\ref{bg1}).

It is important to note that gauge superfield $v$ remains unchanged
when we change the chiral superfields as in (\ref{2chov2}). Thus,
though these transformations have the non-local form the Jacobian of
this change of fields is equal to unit.

The action (\ref{2SCS3}) shows that the chiral superfields have
conventional free propagators, \be i\langle q_+(z) \bar q_+(z')
\rangle =i\langle q_-(z) \bar q_-(z') \rangle   = G_{+-}(z,z')\,,
\ee while the Green's function for the quantum $v$-superfield obeys
\be (H+M-\Delta)G(z,z')=-\delta^7(z-z')\,, \label{G-eq} \ee where we
introduced the notations \bea H &=& \f{ik}{4\pi}\bigg(\bar D^\a D_\a
+\f{\a}{2}(\bar D^2 +
D^2)\bigg) \la{2H}\,, \\
\Delta &=&  M \f{\{\bar D^2, D^2\}}{16\square}\,. \label{Delta} \eea
In Appendix \ref{App-prop} we demonstrate that the solution of
(\ref{G-eq}) can be represented in the form \be
 G(z,z') = \left[\f{i\pi}{k}\f{\bar D^\a D_\a}{\square+\f{4\pi^2 M^2}{k^2}}
-\f{\pi^2 M}{k^2}\f{(\bar D^\a D_\a)^2}{\square(\square+\f{4\pi^2
M^2}{k^2})}
 +\f{i\pi}{2k\a}\f{\bar D^2+ D^2}{\square}
\right]\d^7(z-z')\,. \la{1G} \ee

As we have just shown, the change of fields (\ref{2chov2})
eliminates the mixed propagators among the gauge and matter
superfields. The price for this is a complication of the part of the
action responsible for the interaction vertices. Indeed, after the
change of fields (\ref{2chov2}) the action (\ref{Sint}) acquires
many new vertices which have non-local form
 \bea
 S_{\rm int} =  &-&\int d^7 z \bigg(2\bar Q\, q_+ v^2 +
   2 Q\, \bar q_+ v^2 + 2\bar Q\, q_- v^2 + 2 Q\, \bar q_- v^2 \nn \\
   && +\,2 \bar q_+ q_+ v - 2 \bar q_-  q_- v
   + 2\bar q_+ q_+ v^2 + 2 \bar q_-  q_- v^2   + \f43 \bar Q Q\, v^4 \bigg) \nn\\
  &+&\int d^7z\,d^7z' \bigg(
  4\bar Q_+ G_{-+}(z,z')\, q_+(z)\, v(z)\, v(z')
  + 4 Q_+ G_{+-}(z,z')\, \bar q_+(z)\, v(z)\, v(z') \nn \\
  && +\, 4\bar Q_- G_{-+}(z,z')\, q_-(z)\, v(z)\, v(z')
  + 4 Q_- G_{+-}(z,z')\, \bar q_-(z)\, v(z)\, v(z') \bigg) \nn \\
 &+& 16\int d^7z\,d^7z'\,d^7z'' \, \bar Q Q
 G_{+-}(z,z')\,G_{-+}(z,z'')\, v^2(z)\,v(z')\,v(z'')
 +\ldots
 \la{vertex}
 \eea
Here dots stand for several more terms which have the structure
$q_\pm v^3 $ and $\bar q_\pm v^3$. We omit these terms as the
corresponding vertices cannot appear in the two-loop Feynman
diagrams since we have no mixed $\langle q v \rangle$ and $\langle
\bar q v\rangle$ propagators. We emphasize that the expression
(\ref{vertex}) is a result of identical transformation in local
field theory.

\section{One-loop effective K\"ahler  potential}
The action (\ref{2SCS3}) specifies the one-loop quantum corrections
to the effective action. The (anti)chiral superfields are free and
do not contribute. Thus, the one-loop effective action is given by
the trace of the logarithm of the operator of quadratic fluctuations
for the superfield $v$ \be \Gamma^{(1)} = \f{i}{2} \Tr \ln
(H+M-\Delta)\,, \la{TrLn} \ee where $M$ is the effective mass given
by (\ref{M}) and the operators $H$ and $\Delta$ are defined in
(\ref{2H}) and (\ref{Delta}). It is convenient to represent
(\ref{TrLn}) as a sum of two terms \be \Gamma^{(1)} =\f{i}{2} \Tr
\ln (H+M)+ \f{i}{2} \Tr \ln (1-(H+M)^{-1}\Delta)\,, \la{TrLn1} \ee
and compute them separately.

In the first term in \eq{TrLn1} we expand the logarithm in a series
\bea
 \f{i}{2}\Tr \ln(H+M) &=&  \f{i}{2}\int d^7z\, d^7z' \,
\d^7(z-z')\sum_{n=1}^\infty\f{(-1)^{n-1}}{n M^n}H^n \d^7(z'-z)
\label{4.3}
 \eea
and for the terms in this series we apply the following identities
\begin{subequations}
\label{4.4} \bea
 (\bar D^\a D_\a)^n &=&
 \bigg\{ \begin{array}{ll}
          (4\square)^{k-1} \bar D^\a D_\a & n=2k-1\\
          (4\square)^{k-1} (\bar D^\a D_\a)^2 \qquad\qquad&n=2k
          \end{array} \,,
          \\
 (\bar D^2+D^2)^n &=& \bigg\{ \begin{array}{ll}
               (16\square)^{k-1} (\bar D^2+D^2)&n=2k-1\\
          (16\square)^{k-1} \{\bar D^2,D^2\} \qquad\qquad &n=2k
        \end{array}\,.
\eea
\end{subequations}
Only the terms with four covariant spinor derivatives give
non-trivial rises owing to the standard identity \be
\delta^4(\theta-\theta')\frac1{16}D^2 \bar D^2
\delta^7(z-z')=\delta^7(z-z')\,. \label{id} \ee Then, the expression
(\ref{4.3}) gets local form in the Grassmann variables \bea
 \f{i}{2}\Tr \ln(H+M)
&=& -\f{i}{2}\int d^3 x d^3x' d^4\theta\, \delta^3(x-x')
\f1\square\ln\bigg(\square+\f{4\pi^2 M^2}{k^2} \bigg)\delta^3(x-x') \nn \\
&& +\f{i}{2}\int d^3 x d^3x' d^4\theta\, \delta^3(x-x')
\f1\square\ln\bigg(\square+\f{4\pi^2 M^2}{\a^2 k^2}
\bigg)\delta^3(x-x')\,. \la{trH}
 \eea
Next, we make the Fourier transform for the delta-functions and
compute the momentum integrals
 \be \int
\frac{d^3p}{(2\pi)^3}\frac1{p^2}\ln\left(1-\frac{m^2}{p^2}\right)
=-\frac i{2\pi}|m|\,, \label{mom-int}
 \ee
and get the following result for the term (\ref{4.3})
 \be
 \f{i}{2}\Tr \ln(H+M) =\f2k \bigg(1-\f1\a\bigg) \int d^7z\,\bar Q Q\,. \la{K1_1}
 \ee

To evaluate the second term in \eq{TrLn1} we introduce the Green's
function ${\cal G}(z,z')$ of the operator $H+M$ in \eq{2H} \bea
 (H+M){\cal G}(z,z') &=& - \d^7(z-z')\,, \\
 {\cal G}(z,z') &=& \bigg[\f{i\pi}{k}\f{\bar D^\a D_\a}{\square+\f{4\pi^2 M^2}{k^2}}
 +\f{i\pi}{2k\a}\f{\bar D^2+ D^2}{\square+\f{4\pi^2 M^2}{k^2 \a^2}}
 \nn\\
&& +\f{\pi^2 M}{2k^2}\f{D^\a \bar D^2
D_\a}{\square(\square+\f{4\pi^2 M^2}{k^2})} -\f{\pi^2
M}{4k^2\a^2}\f{\{\bar D^2,D^2\}}{\square(\square+\f{4\pi^2 M^2}{k^2
\a^2})}\bigg]\d^7(z-z')\,. \la{(H+M)}
 \eea
This allows us to represent the second term in \eq{TrLn1} as \bea
 \f{i}{2} \Tr \ln(1-(H+M)^{-1}\Delta)&=& \f{i}{2} \Tr \ln(1+{\cal G}\Delta)
 \nn \\
&=& \f{i}{2} \Tr \ln\bigg(1+A\f{\bar D^2+ D^2}{4}+B\f{\{\bar D^2,
D^2\}}{16}\bigg)\,, \label{4.12} \eea where \be A= \f{2\pi i M}{k
\a} \f{1}{(\square + \f{4\pi^2 M^2}{k^2\a^2})}\,, \qquad B = -
\f{4\pi^2 M^2}{k^2 \a^2} \f{1}{\square(\square + \f{4\pi^2
M^2}{k^2\a^2})}\,. \ee

The argument of the log function in (\ref{4.12}) can be represented
as a product of two factors
 \bea
1+A\f{\bar D^2+ D^2}{16}+B\f{\{\bar D^2, D^2\}}{16} &=&
 \bigg(1+\f{A}{1+\square B}\f{\bar D^2+ D^2}{4}\bigg)
 \bigg(1+ B \f{\{\bar D^2, D^2\}}{16}\bigg)\,.
 \eea
Hence, we have the sum of two terms \be
 \f{i}{2} \Tr \ln(1-(H+M)^{-1}\Delta)=
  \f{i}{2}\Tr \ln \bigg(1+\f{A}{1+\square B}\f{\bar D^2+ D^2}{4}\bigg)+
\f{i}{2} \Tr \ln \bigg(1+ B \f{\{\bar D^2, D^2\}}{16}\bigg)\,. \ee
We expand these log functions in series and in each term apply the
identities (\ref{4.4}). As a result, we get \be
 \f{i}{2} \Tr \ln(1-(H+M)^{-1}\Delta)
= \f{i}{2}\int d^3 x d^3x' d^4\theta\, \delta^3(x-x')
\f1\square\ln\bigg(\f{\square}{\square+\f{4\pi^2 M^2}{\a^2 k^2}}
\bigg)\delta^3(x-x')\,. \ee This expression leads to the same
momentum integral (\ref{mom-int}). Hence, we conclude \be
 \f{i}{2} \Tr \ln(1-(H+M)^{-1}\Delta)=  \f{2}{k\a}\int d^7 z\,
 \bar Q Q\,. \la{K1_2}
\ee

As a result, the one-loop effective action is given by the sum of
(\ref{K1_1}) and (\ref{K1_2})
\begin{subequations}
\bea
\Gamma^{(1)} &=& \int d^7z\, K^{(1)}\,,\\
 \la{Keff1}
K^{(1)}&=& - \f2k\,\bar Q Q\,.\label{4.17b}
 \eea
\end{subequations}
As expected, the one-loop effective K\"ahler potential (\ref{Keff1})
does not contain ultraviolet divergences and is independent of the
gauge-fixing parameter $\a$.

\section{Two-loop effective action}
\label{Sect5}

It is well known that the two-loop quantum contributions to the
effective action are usually represented by the Feynman graphs
having two different topologies which we call ``$\Theta$'' and
``$\infty$''. These diagrams involve cubic and quartic vertices
originating from the action (\ref{vertex}). The lines in these
diagrams correspond to either chiral or gauge superfield propagators
given by (\ref{G+-def}) and (\ref{1G}), respectively. In this
section we compute those two-loop Feynman graphs which contribute to
the effective K\"ahler potential, starting with the diagrams of
topology ``$\infty$''.

\subsection{Graphs with quartic vertices}
\begin{figure}[tb]
\begin{center}
\begin{picture}(240,80)(0,0)
\CArc(25,35)(25,0,180) \CArc(25,35)(25,180,360)
\PhotonArc(75,35)(25,0,180){2}{8}
\PhotonArc(75,35)(25,180,360){2}{8} \Vertex(50,35){2}
\Text(50,5)[]{$\Gamma_{A_1}$}
\PhotonArc(180,35)(25,0,360){2}{18}
\PhotonArc(230,35)(25,0,360){2}{18} \Vertex(205,35){2}
\Text(205,5)[]{$\Gamma_{A_{2a}}$}
\end{picture}

\begin{picture}(240,80)(0,0)
\CArc(25,35)(25,0,90) \Vertex(25,10){2} \CArc(25,35)(25,270,360)
\Vertex(25,60){2} \PhotonArc(25,35)(25,90,270){2}{8}
\PhotonArc(75,35)(25,0,180){2}{8}
\PhotonArc(75,35)(25,180,360){2}{8} \Vertex(50,35){2}
\Text(50,5)[]{$\Gamma_{A_{2b}}$}
\PhotonArc(180,35)(25,0,180){2}{9} \CArc(180,35)(25,180,360)
\Vertex(155,35){2} \CArc(230,35)(25,0,180)
\PhotonArc(230,35)(25,180,360){2}{9} \Vertex(255,35){2}
\Vertex(205,35){2} \Text(205,5)[]{$\Gamma_{A_{2c}}$}
\end{picture}
\caption{Two-loop Feynman graphs with quartic vertices.\label{Fig1}}
\end{center}
\end{figure}
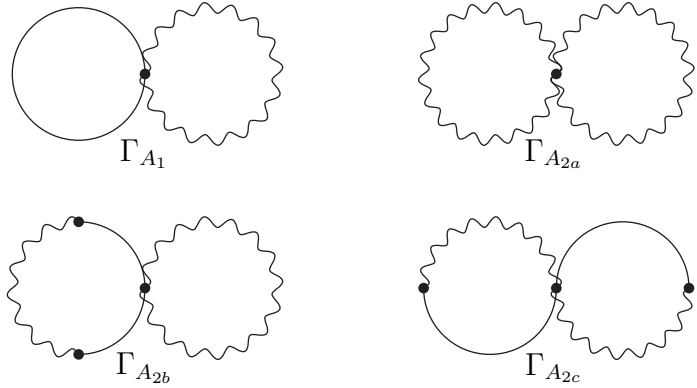

The action (\ref{vertex}) contains two types of quartic vertices:
one vertex of $v^4$ type and the other one of $\bar q q v^2$ type.
Correspondingly, there are two types of two-loop quantum
contributions to the effective action with these vertices \bea
 \Gamma_{A_1} &=& -\int d^5 z_1\, d^5 \bar z_2\, d^7 z_3\, d^7 z_4\,
\f{\d^4 S}{\d v(z_4) \d v(z_3) \d\bar q(z_2) \d q(z_1)}
\, G_{+-}(z_1,z_2)\,G(z_3,z_4)\,,\la{A1} \\
 \Gamma_{A_2}
 &=& -\f1{8}\int d^7 z_1\, d^7 z_2\, d^7 z_3\, d^7 z_4\,
\f{\d^4 S}{\d v(z_4) \d v(z_3)\d v(z_2)\d v(z_1)}
G(z_1,z_2)\,G(z_3,z_4)\,. \la{A2} \eea Note that the vertex in
(\ref{A1}) has simple local form \bea
 \f{\d^4 S}{\d v(z_4) \d v(z_3) \d\bar q(z_2) \d q(z_1)}
 = -4  \bigg(-\f{\bar D^2_1}{4}\bigg)
  \bigg(\d_-(z_1-z_2)\d^7(z_1-z_3) \d^7(z_1-z_4)\bigg)\,.
\eea Using this expression we restore the full superspace measure in
(\ref{A1}), $d^7z_1 = -\frac14 \bar D^2_1 d^5 z_1$ and integrate
over $z_2$, $z_3$ and $z_4$ using the delta-functions \be
 \Gamma_{A_1} = 4 \int d^7 z\, G_{+-}(z,z)\,G(z,z)\,.
\ee This contribution can be visualized by the Feynman graph
$\Gamma_{A_1}$ in Fig.\ \ref{Fig1}.
Formally, the propagators (\ref{G+-def}) and (\ref{1G}) have enough
$D$-factors to get a non-trivial result, but the bosonic part of the
propagator $G_{+-}$ vanishes at coincident space-time points in the
frame of dimensional regularization. Thus, this Feynman graph does
not contribute to the effective K\"ahler potential
 \be
 \Gamma_{A_1} = 0\,.
 \ee
We point out that, in general, this Feynman graph is non-trivial,
but it vanishes on the background (\ref{bg1}).

Consider now the contribution (\ref{A2}). It contains the quartic
vertex which has the following form
\begin{multline}
\f{\d^4 S}{\d v(z_4) \d v(z_3)\d v(z_2)\d v(z_1)} =
 - 32\, \bar Q Q\, \d^7(z_1-z_2)\d^7(z_1-z_3)\d^7(z_1-z_4)   \\
  -32 \bar Q Q \bigg(
    G_{+-}(z_2,z_3) G_{-+}(z_2,z_1)\d^7(z_2-z_4)
  + G_{+-}(z_2,z_4) G_{-+}(z_2,z_1)\d^7(z_2-z_3) \\
  \,\quad\qquad + G_{+-}(z_3,z_2) G_{-+}(z_3,z_1)\d^7(z_3-z_4)
  + G_{+-}(z_3,z_1) G_{-+}(z_3,z_2)\d^7(z_3-z_4) \\
  \,\quad\qquad + G_{+-}(z_2,z_1) G_{-+}(z_2,z_3)\d^7(z_2-z_4)
  + G_{+-}(z_2,z_1) G_{-+}(z_2,z_4)\d^7(z_2-z_3) \\
  \,\quad\qquad + G_{+-}(z_1,z_2) G_{-+}(z_1,z_3)\d^7(z_1-z_4)
  + G_{+-}(z_1,z_2) G_{-+}(z_1,z_4)\d^7(z_1-z_3) \\
  \,\quad\qquad + G_{+-}(z_1,z_3) G_{-+}(z_1,z_2)\d^7(z_1-z_4)
  + G_{+-}(z_1,z_4) G_{-+}(z_1,z_2)\d^7(z_1-z_3) \\
  \,\qquad\qquad + G_{+-}(z_1,z_4) G_{-+}(z_1,z_3)\d^7(z_1-z_2)
  + G_{+-}(z_1,z_3) G_{-+}(z_1,z_4)\d^7(z_1-z_2) \bigg)\,.
      \la{Vert5}
\end{multline}
The term in the first line corresponds to the local part of this
vertex while the other terms are non-local since they involve the
Green's functions $G_{+-}$ and $G_{-+}$. The contribution from the
local part of this vertex is represented by the Feynman graph
$\Gamma_{A_{2a}}$ in Fig.\ \ref{Fig1}. The other non-local terms in
(\ref{Vert5}) correspond to $\Gamma_{A_{2b}}$ and $\Gamma_{A_{2c}}$.

Let us consider first the Feynman graph $\Gamma_{A_{2a}}$
corresponding to the local part of the vertex (\ref{Vert5}). Using
the delta-functions in this vertex we integrate over $dz_2$, $dz_3$
and $dz_4$ \be
 \Gamma_{A_{2a}}
 =  4 \int d^7 z\, \bar Q Q\,G(z,z)\,G(z,z)\,. \la{A2_a}
\ee To compute this expression we have to consider the Green's
function of the gauge superfield (\ref{1G}) at coincident superspace
points. The details of these computations are collected in Appendix
\ref{AppA1}. The result is \be \Gamma_{A_{2a}} = - \f{4}{k^2} \int
d^7 z \, \bar Q Q\,. \label{5.8} \ee

The Feynman graphs $\Gamma_{A_{2b}}$ and $\Gamma_{A_{2c}}$ in
 Fig.\ \ref{Fig1} correspond to the following analytic expressions
 \bea
\Gamma_{A_{2b}}
 &=&  16 \int d^7 z_1 d^7 z_2 d^7 z_3 \, \bar Q Q\,
 G_{+-}(z_1,z_3)G_{-+}(z_2,z_3) G(z_1,z_2)\,G(z_3,z_3) =0\,, \la{A2_b}
 \\
\Gamma_{A_{2c}}
 &=&  16 \int d^7 z_1 d^7 z_2 d^7 z_3 \, \bar Q Q\,
 G_{+-}(z_3,z_2)G_{-+}(z_1,z_2)\, G(z_1,z_2) G(z_3,z_2) =0\,. \la{A2_c}
 \eea
To evaluate these expressions we have to integrate by parts the
covariant spinor derivatives which are present in the propagators
$G_{+-}$ given in (\ref{G+-def}). It is possible to distribute the
derivatives in such a way that the operator $\bar D^2 D^2$ acts on
the propagator (\ref{1G}). Then, it is easy to see that both
contributions (\ref{A2_b}), (\ref{A2_c}) vanish owing to the
identity \eq{G1=0} \be \Gamma_{A_{2b}}=\Gamma_{A_{2c}}=0\,. \ee

We conclude that the Feynman graphs represented in Fig.\ \ref{Fig1}
give rise to the effective action (\ref{5.8}). The corresponding
part of the effective K\"ahler potential has a form similar to
(\ref{4.17b}).

\subsection{Graphs with cubic vertices}

According to the action (\ref{vertex}), there are the following
three types of cubic vertices \bea
 \f{\d^3 S_{\rm int}}{\d v(z_3) \d \bar q_\pm(z_2) \d q_\pm (z_1)} &=&
  \pm (-2)  \bigg( -\f{ \bar D^2_1}{4}\bigg)
  \bigg(\d_-(z_1-z_2)\d^7(z_1-z_3)\bigg),\label{5.12}\\
\f{\d^3 S_{\rm int}}{\d v(z_3) \d v(z_2) \d q_\pm (z_1)} &=&
   -4 \bar Q_\pm \bigg( -\f{\bar D^2_1}{4}\bigg)
  \bigg(\d^7(z_1-z_2)\d^7(z_1-z_3)\bigg)\la{Vert1} \\
  && +\, 4 \bar Q_\pm \bigg( -\f{\bar D^2_1}{4}\bigg)
 \bigg(G_{-+}(z_1,z_2)\d^7(z_1-z_3)+ G_{-+}(z_1,z_3)
 \d^7(z_1-z_2)\bigg),\nn \\
 \f{\d^3 S_{\rm int}}{\d v(z_3) \d v(z_2) \d \bar q_\pm(z_1)} &=&
  -4  Q_\pm \bigg( -\f{ D^2_1}{4}\bigg)
    \bigg(\d^7(z_1-z_2)\d^7(z_1-z_3)\bigg) \label{Vert1_}\\
  && +\, 4 Q_\pm \bigg( -\f{ D^2_1}{4}\bigg)
 \bigg(G_{+-}(z_1,z_2)\d^7(z_1-z_3)+ G_{+-}(z_1,z_3)
 \d^7(z_1-z_2)\bigg).\nn
 \eea
Using these vertices and the propagators (\ref{G+-def}), (\ref{1G})
it is possible to construct the following two types of two-loop
contributions to the effective action \bea \Gamma_{B_1} &=& - \int
d^5 z_1\, d^5 \bar z_2\, d^7 z_3\, d^5 z_4\, d^5 \bar z_5\,d^7 z_6\,
\f{\d^3 S}{\d v(z_3) \d \bar q(z_2) \d q(z_1)} \cdot \f{\d^3 S}{\d
v(z_6) \d \bar q(z_5) \d q (z_6)}  \nn
\\
&& \qquad  \times\, G_{+-}(z_1,z_5) G_{+-}(z_4,z_2) G(z_3,z_6)\,,
\la{B1}
\\
\Gamma_{B_2} &=& - \f12 \int d^5 z_1\, d^7 z_2\, d^7 z_3\, d^5 \bar
z_4\, d^7 z_5\,d^7 z_6\, \f{\d^3 S}{\d v(z_3) \d v(z_2) \d q(z_1)}
\cdot \f{\d^3 S}{\d v(z_6) \d v(z_5) \d \bar q(z_4)} \nn
\\
&& \qquad \times \bigg( G_{+-}(z_1,z_4) G(z_2,z_5) G(z_3,z_6) +
G_{+-}(z_1,z_4) G(z_2,z_6) G(z_3,z_5) \bigg). \la{B2} \eea Consider
them separately.

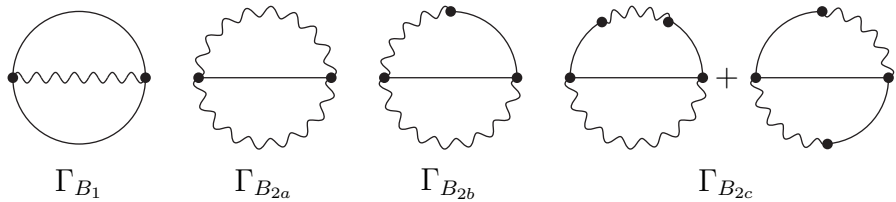
\begin{figure}[tbh]
\begin{center}
\begin{picture}(340,90)(0,0)
\CArc(25,45)(25,0,180) \Photon(0,45)(50,45){2}{7}
\CArc(25,45)(25,180,360) \Vertex(50,45){2} \Vertex(0,45){2}
\Text(25,5)[]{$\Gamma_{B_1}$}
\PhotonArc(95,45)(25,0,180){2}{9} \Line(70,45)(120,45)
\PhotonArc(95,45)(25,180,360){2}{9} \Vertex(70,45){2}
\Vertex(120,45){2} \Text(95,5)[]{$\Gamma_{B_{2a}}$}
\PhotonArc(165,45)(25,90,180){2}{5} \CArc(165,45)(25,0,90)
\Line(140,45)(190,45) \PhotonArc(165,45)(25,180,360){2}{9}
\Vertex(140,45){2} \Vertex(165,70){2} \Vertex(190,45){2}
\Text(165,5)[]{$\Gamma_{B_{2b}}$}
\CArc(235,45)(25,0,60) \CArc(235,45)(25,120,180)
\PhotonArc(235,45)(25,60,120){2}{4} \Line(210,45)(260,45)
\PhotonArc(235,45)(25,180,360){2}{9} \Vertex(210,45){2}
\Vertex(222,66){2} \Vertex(247,66){2} \Vertex(260,45){2}
\CArc(305,45)(25,90,180) \PhotonArc(305,45)(25,0,90){2}{5}
\Line(280,45)(330,45) \PhotonArc(305,45)(25,180,270){2}{5}
\CArc(305,45)(25,270,360) \Vertex(280,45){2} \Vertex(305,70){2}
\Vertex(307,20){2} \Vertex(330,45){2} \Text(270,45)[]{+}
\Text(270,5)[]{$\Gamma_{B_{2c}}$}
\end{picture}
\end{center}
\caption{Two-loop Feynman graphs with cubic vertices.\label{Fig2}}
\end{figure}

Using the $\bar D^2$ operators in the vertex (\ref{5.12}) it is
possible to restore full superspace measure in some of the integrals
in (\ref{B1}) and to perform these integrals using the
delta-functions. Then, we get the following representation for
(\ref{B1}) \be
  \Gamma_{B_1} = -4 \int d^7 z_1\,  d^7 z_2\,
    G_{+-}(z_1,z_2) G_{-+}(z_1,z_2) G(z_1,z_2)\,.\label{5.17}
\ee This analytic expression is represented by the Feynman graph
$\Gamma_{B_1}$ in Fig.\ \ref{Fig2}.
Next, we use the derivatives $D^2$ and $\bar D^2$ in the propagator
$G_{+-}(z_1,z_2)$ and integrate them by parts such that the
identities \eq{prG1}, \eq{prG2} and \eq{G1=0} can be used. It is
easy to see that owing to these properties of the propagator
$G(z,z')$ the contribution (\ref{5.17}) vanishes on the considered
background (\ref{bg1})
 \be
 \Gamma_{B_1} =0\,.
 \ee

Let us consider now the part of effective action (\ref{B2}). Using
the operators $D^2$ and $\bar D^2$ in the vertices (\ref{Vert1}) and
(\ref{Vert1_}) we restore full superspace measures and perform some
of the integrals using the superspace delta-functions. Then, the
contribution (\ref{B2}) can be represented as a sum of the following
terms
 \be \Gamma_{B_2} = \Gamma_{B_{2a}} + \Gamma_{B_{2b}} +
\Gamma_{B_{2c}}\,,
 \ee
 where
 \bea
\Gamma_{B_{2a}}
&=& -16 \int d^7 z_1\,d^7 z_2\, \bar Q Q\,G_{+-}(z_1,z_2) G(z_1,z_2) G(z_1,z_2) \,, \\
\Gamma_{B_{2b}}&=& 64\int d^7 z_1\,d^7 z_2\,d^7 z_3\,
    \bar Q Q\, G_{+-}(z_2,z_3)G_{+-}(z_1,z_2) G(z_1,z_3) G(z_1,z_2)\,, \\
\Gamma_{B_{2c}}&=& - 32\int d^7 z_1\,d^7 z_2\,d^7 z_3\,d^7 z_4\,
    \bar Q Q\, G_{-+}(z_1,z_2)G_{+-}(z_3,z_4) G_{+-}(z_1,z_3)\nn \\
&& \times\bigg( G(z_1,z_3) G(z_2,z_4) + G(z_1,z_4) G(z_2,z_3)
\bigg)\,.
 \eea
These contributions to the effective action correspond to the
Feynman graphs $\Gamma_{B_{2a}}$, $\Gamma_{B_{2b}}$ and
$\Gamma_{B_{2c}}$ in Fig.\ \ref{Fig2}. The details of the
computations of these diagrams are given in Appendix A.2. Only the
final results are written down here:
\begin{subequations}
\label{5.23} \bea
 \Gamma_{B_{2a}} &=& - \f{4}{k^2} \int d^7 z\,\, \bar Q Q\,
\bigg(\f{1}{\ep} - \gamma - 2\ln \f{\bar Q Q}{k \mu}\bigg)
+ \f{8\ln2}{k^2} \int d^7 z\,\bar Q Q \nn \\
&&+\f{4}{k^2 \a^2} \int d^7 z\,
   \bar Q(z)\bigg(\f{1}{\ep}-\gamma\bigg) Q(z)\,,\\
 \Gamma_{B_{2b}}&=&
-\f{8}{k^2 \a^2} \int d^7 z\,
   \bar Q(z)\bigg(\f{1}{\ep}-\gamma \bigg) Q(z) \nn\\
&& - \f{8}{k^2} \int d^7 z\,\, \bar Q Q\, \bigg(\f{1}{\ep} - \gamma
- 2\ln \f{\bar Q Q}{k \mu}\bigg)
 - \f{4(1-2\ln2)}{k^2} \int  d^7 z\,\bar Q Q \,,\\
 \Gamma_{B_{2c}}&=&\f{4}{k^2 \a^2} \int d^7 z\,
   \bar Q(z)\bigg(\f{1}{\ep}-\gamma \bigg) Q(z)\,.
\eea
\end{subequations}
Here $\varepsilon$ is the parameter of dimensional regularization,
$d=3-2\varepsilon$, $\varepsilon\to 0$, and $\gamma$ is the Euler
constant. Let us collect the divergent and finite terms in
(\ref{5.23}) separately \bea
\Gamma_{B_2} &=& \Gamma_{B_2,\rm div} + \Gamma_{B_2, \rm fin}\,,\label{5.24}\\
\Gamma_{B_2,\rm div} &= & -\frac{12}{\varepsilon k^2}
\int d^7z \, \bar Q Q\,, \label{5.25} \\
\Gamma_{B_2, \rm fin} &=& \int d^7z\, \bar Q Q \left( \frac{16\ln 2
- 4 +12\gamma}{k^2} +\frac{24}{k^2} \ln \frac{\bar QQ}{k\mu}
\right). \label{5.26}
 \eea
In is important to note that all terms containing the gauge-fixing
parameter $\alpha$ cancel each other out in (\ref{5.23}) and the
final result (\ref{5.24}) is $\alpha$-independent. This result is
expected since we have taken the gauge-independent part of the
effective action into account.

The divergent contribution to the effective action (\ref{5.25}) has
the structure of the classical action for the chiral superfield. It
can be eliminated by adding the corresponding counterterm to the
bare action (\ref{2CS0}). Further we concentrate only on the finite
terms which contribute to the effective K\"ahler potential.

\subsection{Two-loop effective K\"ahler potential}
The sum of two-loop finite contributions to the effective action
(\ref{5.8}) and (\ref{5.26}) can be written as \be \Gamma^{(2)}_{\rm
fin} = -\int d^7z\, K^{(2)}\,, \ee where $K^{(2)}$ is the two-loop
quantum correction to the effective K\"ahler potential \be K^{(2)}=
2 \bar Q Q \bigg( \f{4}{k^2} - \f{8\ln2}{k^2} - \f{6}{k^2} \g
 - \f{12}{k^2} \bar Q Q\,\ln \f{\bar Q Q}{k \m}\bigg) \,.
\label{5.28} \ee

Let us now consider the full effective K\"ahler potential which
comprises the one-loop (\ref{4.17b}) and two-loop (\ref{5.28})
quantum contributions as well as the classical part
 \be
 K_{\rm eff}(\bar Q, Q) =2 \bar Q Q\bigg( 1 - \f{1}{k} + \f{4}{k^2} -
\f{8\ln2}{k^2} - \f{6\gamma}{k^2} - \f{12}{k^2} \,\ln \f{\bar Q
Q}{k\m} \bigg)\,. \ee
 Here $\mu$ is the normalization point which is
usually fixed from the condition
 \be
 \f{\pr^2 K_{\rm eff}(\bar Q, Q)}{\pr \bar Q \pr Q}\bigg|_{Q=Q_0} =
 2\,.
\ee With such a normalization we get the final expression for the
effective K\"ahler potential in the two-loop approximation \be
K_{\rm eff}= 2\, \bar Q Q - \f{24}{k^2}\, \bar Q Q \bigg( \ln\f{\bar
Q\, Q}{\bar Q_0 Q_0}-2\bigg)\,. \la{Keff2} \ee

The effective K\"ahler potential (\ref{Keff2}) deserves the
following comments:
\begin{itemize}
\item The form of the effective K\"ahler potential is very similar
to the one-loop effective K\"ahler potential in four-dimensional
$\cN=1$ supersymmetric gauge theories interacting with chiral matter
which were studied in \cite{BKY,deWit1996,PW,Grisaru96}. In fact,
this form is universal in the sense that it is dictated by
logarithmic quantum divergences which appear in one loop in four
dimensions and start from two loops in three-dimensional model under
considerations.
\item We did all the quantum computations keeping the
gauge-fixing parameter $\alpha$ arbitrary, but found that the
effective K\"ahler potential (\ref{Keff2}) is independent of
$\alpha$. This is a manifestation of the fact that we computed the
low-energy effective action on the background (\ref{bg1}) which
solves not only the classical, but also effective quantum equations
of motion in the one-loop approximations.
\item Let us consider the model of $\cN=2$ supersymmetric
electrodynamics (\ref{2CS0}) which is known to be superconformal at
the classical level. The effective K\"ahler potential explicitly
breaks the superconformal invariance since it involves dimensional
parameters $Q_0$ and $\bar Q_0$. Thus, the superconformal invariance
is broken by two-loop quantum corrections.
\end{itemize}

\section{Conclusions}

In the present paper, we have computed the two-loop effective
K\"ahler potential in the $\cN=2$ SQED with Chern-Simons kinetic
term for the gauge superfield. The result (\ref{Keff2}) resembles
the four-dimensional one-loop effective K\"ahler potential
\cite{BKY,deWit1996,PW,Grisaru96} since its form is stipulated by
logarithmic quantum divergences.

The calculations have been done in the framework of the background
field method with a one-parametric family of gauges. It was proven
that the resultant effective K\"ahler potential is gauge
independent. Also, we want to emphasize that we have used the
"non-standard" change of variables in the functional integral to
diagonalize the propagator. Such a procedure creates the new
non-local interaction vertices in the supergraphs. We have shown
that these new vertices do not lead to obstacles in computations. In
the conclusions, let us discuss the possible future development of
the obtained results.

The most obvious generalization is to consider non-Abelian $\cN=2$
Chern-Simons matter theories. We expect that the form of the
effective K\"ahler potential in these theories should be similar to
(\ref{Keff2}), but many more quantum computations are required since
one has to take into account more Feynman graphs in non-Abelian
theories including, in particular, ghost field contributions. We
will leave these issues for further studies.

It is possible to include more parameters in the classical action
such as the masses of chiral matter superfields and the Yang-Mills
gauge coupling. It is interesting to study how the effective
K\"ahler potential depends on the values of all these parameters.

The Chern-Simons-matter theories with $\cN>2$ supersymmetry are
known to be UV-finite \cite{Avdeev1,Avdeev2,BILPSZ}. Hence, the
effective K\"ahler potential in these models can receive only finite
quantum corrections. For $\cN=4$ supersymmetric models there is a
non-renormalization theorem \cite{Argyres,IntSeib} which forbids
perturbative quantum corrections to the moduli space in the Higgs
branch described by the effective K\"ahler potential. However, it is
not clear whether this applies to $\cN=3$ supersymmetric
gauge-matter theories. Therefore, it would be interesting to study a
structure of the effective K\"ahler potential in the $\cN=3$ gauge
theory.

\vspace{30pt} \noindent
{\bf Acknowledgments}\\[3mm]
The authors are extremely grateful to I.B. Samsonov for his
collaboration during the earlier stages of work and valuable
discussions. The authors are thankful to S.J. Tyler for
correspondence. The work was supported in part by the RFBR grant Nr.
15-02-03594, by LRSS grant Nr. 88.2014.2 and DFG grant LE 838/12-2.
I.L.B. acknowledges the support from RFBR grants Nr. 15-02-06670.
The work of B.S.M. was partly supported by the RFBR grant for young
researchers Nr. 14-02-31201. I.L.B. and B.S.M. are thankful to the
grant of Russian Ministry of Education and Science, project
2014/387/122 for partial support.

\appendix

\section{Gauge superfield propagator}
\label{App-prop}

In this Appendix we consider some useful properties of the gauge
superfield propagator introduced in sect.\ \ref{Sect-props}. Recall
that, after gauge fixing and performing the non-local change of
fields (\ref{2chov2}), the quadratic action for gauge superfields is
defined by the operators (\ref{2H}) and (\ref{Delta}). Using the
identity
 \bea
 (\bar D^\a D_\a)^2 =  4\square -\f14\{D^2,\bar D^2\} \la{id1}
 \eea
the operator of quadratic fluctuations of the gauge superfield can
be rewritten as
 \bea
H+M-\Delta &=& H + M-M\f{\{D^2,\bar D^2\}}{16\square} =
H+\f{M}{4\square}(\bar D^\a D_\a)^2 \nn \\
 &=&\f{ik}{4\pi}\bar D^\a D_\a +\f{M}{4\square}(\bar D^\a D_\a)^2
 +\f{ik\a}{8\pi}(\bar D^2 + D^2)\,.
 \eea
Then, it is easy to check that the distribution \be
 G(z,z') = \bigg[\f{i\pi}{k}\f{\bar D^\a D_\a}{\square+\f{4\pi^2 M^2}{k^2}}
-\f{\pi^2 M}{k^2}\f{(\bar D^\a D_\a)^2}{\square(\square+\f{4\pi^2
M^2}{k^2})}
 +\f{i\pi}{2k\a}\f{\bar D^2+ D^2}{\square}
\bigg]\d^7(z-z') \la{1G_} \ee solves the equation for the gauge
superfield propagator \be
 (H+M-\Delta) G(z,z') = - \d^7(z-z')\,.
\ee Using (\ref{id1}) the propagator (\ref{1G_}) can be identically
rewritten as \bea
 G(z,z') &=& \bigg[
-\f{4\pi^2 M}{k^2}\f{1}{\square+\f{4\pi^2 M^2}{k^2}}
+\f{i\pi}{k}\f{\bar D^\a D_\a}{\square+\f{4\pi^2 M^2}{k^2}}
+\f{\pi^2 M}{4k^2}\f{\{\bar D^2,D^2\}}{\square(\square+\f{4\pi^2 M^2}{k^2})}\nn\\
 &&+\f{i\pi}{2k\a}\f{\bar D^2+ D^2}{\square}
\bigg]\d^7(z-z')\,. \la{G} \eea It is straightforward to check that
(\ref{G}) has the following useful properties \bea
 \bar D^2 G(z,z') &=& \f{i\pi}{2k\a} \f{\bar D^2 D^2}{\square} \d^7(z-z')
 \,, \la{prG1}\\
 \f{D^2 \bar D^2}{16} G(z,z') &=& \f{i\pi}{2k\a} D^2 \d^7(z-z')
 = - \f{2i\pi}{k\a}\, \d_-(z,z')  \,, \la{prG2}
\eea where $\d_-(z,z')$ is the antichiral delta function.

In loop quantum computations, we need the expressions for the gauge
superfield propagator (\ref{G}) and its derivatives at coincident
Grassmann coordinates \bea
 G(z,z')\bigg|_{\t=\t'} &=& \f{8\pi^2 M}{k^2}
 \f{1}{\square(\square+\f{4\pi^2 M^2}{k^2})} \d^3(x-x')\,, \la{G=0}
\\
\bar D^2 D^2 G(z,z')\bigg|_{\t=\t'} &=& 0\,,  \la{G1=0}\\
 \bar D^2 G(z,z')=D^2 G(z,z')\bigg|_{\t=\t'} &=&
\f{8i\pi}{k\a}  \f{1}{\square} \d^3(x-x')\,, \la{G2=0}
\\
\bar D^\a D_\a \, G(z,z') \bigg|_{\t=\t'} &=& - \f{8 i \pi}{k}
\f{1}{\square+\f{4\pi^2 M^2}{k^2}}\, \d^3(x-x')\,, \la{G3=0} \\
D_\b\bar D_\a G(z,z')\bigg|_{\t=\t'} &=& \f{4i\pi
}{k}\f{1}{\square+\f{4\pi^2 M^2}{k^2}}\bigg[\ep_{\a\b}
- \f{2\pi M }{k}\f{\pr_{\a\b}}{\square}\bigg]\d^3(x-x') \,, \la{G4=0} \\
\f{ D^2 \bar D^2}{16} D_\b\bar D_\a G(z,z')\bigg|_{\t=\t'}
&=&\f{4i\pi }{k} \f{\square}{\square+\f{4\pi^2
M^2}{k^2}}\bigg[\ep_{\a\b} - \f{2\pi M
}{k}\f{\pr_{\a\b}}{\square}\bigg]\d^3(x-x') \,. \la{G5=0}
 \eea

\section{Some details of computations of two-loop diagrams}
In this Appendix we collect some details of quantum computations of
two-loop Feynman graphs which were considered in sect.\ \ref{Sect5}.
We start by considering the graphs in Fig.\ \ref{Fig1}.
\subsection{Diagram $\Gamma_{A_2}$ }
\label{AppA1} The Feynman graph  $\Gamma_{A_{2a}}$ in Fig.\
\ref{Fig1} contains two gauge superfield propagators which meet at
one quartic vertex. We will use this propagator in the form \eq{G}.
Using the superspace delta-function which is present in this
propagator we integrate over one set of Grassmann variables $\t'$
\bea
  \Gamma_{A_{2a}}
 &=&  4 \int d^7 z\,d^7 z'\,\d^7(z-z')  \bar Q Q\,G(z,z')\,G(z,z')  \nn\\
 &=& \f{4\cdot64\pi^4 }{k^4}  \int d^7 z\,d^3 x'\,\d^3(x-x')  \bar Q Q \, M^2
 \nn\\&&\times
 \f{1}{\square(\square+\f{4\pi^2 M^2}{k^2})} \d^3(x-x')\,
  \f{1}{\square(\square+\f{4\pi^2 M^2}{k^2})} \d^3(x-x')\,.
  \label{B1_}
\eea Here we used the identity \eq{G1=0}. For bosonic
delta-functions in (\ref{B1_}) we perform the Fourier transform and
compute the resulting momentum integral using \eq{J}
 \bea
\Gamma_{A_{2a}} &=& \f{4\cdot64\pi^4 }{k^4} \int d^7 z\, \bar Q Q \,
M^2
\bigg(\int\f{d^3p}{(2\pi)^3} \f{1}{p^2(p^2-\f{4\pi^2 M^2}{k^2})}\bigg)^2 \nn \\
&=& - \f{4 }{k^2} \int d^7 z\, \bar Q Q \,.
 \eea
This contribution to the effective action is finite and has the form
of classical action for a free chiral superfield $Q$.

\subsection{Diagrams $\Gamma_{B_2}$ }
Consider now the details of quantum computations of Feynman graphs
given in Fig.\ \ref{Fig2}.
\subsubsection{Diagram $\Gamma_{B_{2a}}$}
The diagram $\Gamma_{B_{2a}}$ in Fig.\ \ref{Fig2} contains two gauge
superfield propagators and one (anti)chiral propagator which meet at
two cubic vertices. For the gauge superfield propagator $G(z,z')$ we
will use the representation \eq{G} while the (anti)chiral propagator
is given by (\ref{G+-def})
 \bea
\Gamma_{B_{2a}}
&=& -16 \int d^7 z_1\,d^7 z_2\, \bar Q Q\, G_{+-}(z_1,z_2) G(z_1,z_2) G(z_1,z_2) \\
&=& -16 \int d^7 z_1\,d^7 z_2\, \bar Q Q\,
G_{+-}(z_1,z_2)\,G(z_1,z_2) \nn \\
&\times&  \bigg[ -\f{4\pi^2 M}{k^2}\f{1}{\square+\f{4\pi^2
M^2}{k^2}} +\f{i\pi}{k}\f{\bar D^\a D_\a}{\square+\f{4\pi^2
M^2}{k^2}}
+\f{\pi^2 M}{4k^2}\f{\{\bar D^2,D^2\}}{\square(\square+\f{4\pi^2 M^2}{k^2})}\nn\\
 && +\f{i\pi}{2k\a}\f{\bar D^2+ D^2}{\square}
\bigg]\d^7(z_1-z_2)\,.
 \eea
We rewrite the last integrals as a sum of four terms
 \bea
\Gamma_{B_{2a}} &=& \f{64\pi^2}{k^2}
 \int d^7 z_1 d^7 z_2\,  \bar Q Q M\, G_{+-}(z_1,z_2)\,G(z_1,z_2)
\f{1}{\square+\f{4\pi^2 M^2}{k^2}}\d^7(z_1-z_2) \nn \\
&&-\f{16 i\pi}{k} \int d^7 z_1d^7 z_2\,\bar Q
Q\,G_{+-}(z_1,z_2)\,G(z_1,z_2)
\f{\bar D^\a D_\a}{\square+\f{4\pi^2 M^2}{k^2}}\d^7(z_1-z_2) \nn \\
&&-\f{4\pi^2}{k^2} \int d^7 z_1 d^7 z_2\,  \bar Q Q M\,
G_{+-}(z_1,z_2)\,G(z_1,z_2)
\f{\{\bar D^2,D^2\}}{\square(\square+\f{4\pi^2 M^2}{k^2})}\d^7(z_1-z_2) \nn\\
&& - \f{8i\pi}{k\a}\int d^7 z_1 d^7 z_2\,  \bar Q Q
\,G_{+-}(z_1,z_2)\,G(z_1,z_2) \f{\bar D^2+
D^2}{\square}\d^7(z_1-z_2)\,. \la{A.10}
 \eea

In first line in (\ref{A.10}) we use the explicit form of the full
superspace delta-function
$\d^7(z_1-z_2)=\d^4(\t_1-\t_2)\d^3(x_1-x_2)$ and integrate over
$\t_2$ using \eq{G=0}. In the second line in (\ref{A.10}) we
integrate by parts the derivatives of the $\bar D^\a D_\a$ operator
and then integrate over $\t_2$ using \eq{G3=0}. In the third line in
(\ref{A.10}) we integrate by parts the covariant spinor derivatives
contained in the Green function $G_{+-}(z_1,z_2)$ and integrate over
$\t_2$ using \eq{G=0}. After this, the terms in the first and third
lines cancel against each other.

The term in the second line of (\ref{A.10}) can be rewritten as \be
\Gamma_{B_{2a}}= -\f{128\pi^2}{k^2} \int d^7 z_1\,d^3 x_2\, \bar Q Q
\f{1}{\square}\d^3(x_1-x_2)\, \f{1}{\square+\f{4\pi^2
M^2}{k^2}}\d^3(x_1-x_2) \f{1}{\square+\f{4\pi^2
M^2}{k^2}}\d^3(x_1-x_2) \,. \ee Passing to the momentum
representation we integrate over space-time variable $x_2$ and
calculate the momentum integrals using \eq{I} and \eq{I2}
 \bea
\Gamma_{B_{2a}}&=&  \f{128 \pi^2}{k^2} \int d^7 z\,\, \bar Q Q\,
   \int \f{d^3 p\, d^3 q}{(2\pi)^6 }\,
       \f{1}{(p+q)^2(p^2-\f{4\pi^2 M^2}{k^2})(q^2-\f{4\pi^2 M^2}{k^2})}  \nn\\
&=&- \f{4}{k^2} \int d^7 z\,\, \bar Q Q\, \bigg(\f{1}{\ep} - \gamma
- 2\ln \f{\bar Q Q}{k \mu}\bigg) + \f{8\ln2}{k^2} \int d^7 z\,\bar Q
Q\,.      \quad (\ep\to 0) \label{B7_}
 \eea

Let us consider the term in the last line in \eq{A.10}. We integrate
by parts the covariant spinor derivatives which are contained in the
chiral superfield propagator \eq{G+-def} keeping in mind the
identity \eq{G2=0}
 \bea
&& -\f{8i\pi}{k\a}\int d^7 z_1 d^7 z_2\, \bar Q(z_1) Q(z_2)
\,G_{+-}(z_1,z_2)\,G(z_1,z_2)
\f{\bar D^2+ D^2}{\square}\d^7(z_1-z_2)  \nn \\
&=&\f{128\pi^2}{k^2 \a^2} \int d^7 z_1 d^3 x_2\, \bar Q(x_1,\t_1)
Q(x_2,\t_1) \f{1}{\square}\d^3(x_1-x_2) \f{1}{\square}\d^3(x_1-x_2)
\f{1}{\square}\d^3(x_1-x_2)\,.
 \nn
 \eea
Here we also integrated over $\t_2$. Next, we pass to the momentum
representation and integrate over $x_1$ and $x_2$ \bea
-\f{128\pi^2}{k^2 \a^2} \int d^4 \t \int \f{d^3l}{(2\pi)^3}\bar
Q(l,\t) Q(-l,\t) \, \int\f{d^3p d^3q}{(2\pi)^6}\f{1}{p^2 q^2
(p+q+l)^2}\,.
 \eea
This momentum integral can be evaluated using \eq{I3} after the
change of integration variable $p\to p-l$ \bea &&-\f{128\pi^2}{k^2
\a^2} \int d^4 \t \int \f{d^3l}{(2\pi)^3}\bar Q(l,\t) Q(-l,\t)
\int\f{d^3p d^3q}{(2\pi)^6}\f{1}{(p-l)^2 q^2 (p+q)^2} \nn\\
&&= \f{4}{k^2 \a^2} \int d^4 \t \int \f{d^3l}{(2\pi)^3}\bar
Q(l,\t)\bigg(\f{1}{\ep}-\gamma-\ln(- l^2) \bigg) Q(-l,\t)\,,
 \eea
In the coordinate representation this expression reads
  \bea &&\f{4}{k^2 \a^2} \int d^7 z\,
   \bar Q(z)\bigg(\f{1}{\ep}-\gamma-\ln\square \bigg) Q(z)\,. \quad (\ep\to
   0)
 \la{A.12}
 \eea
Note that the last term in \eq{A.12} contains the operator
$\ln\square$ which can be discarded for the constant superfield
background (\ref{bg1}) which we use to compute the effective
K\"ahler potential.

Finally, we combine the two non-trivial contributions (\ref{B7_})
and (\ref{A.12}) and get the following result for the quantum
contributions corresponding to the Feynman graph $\Gamma_{B_{2a}}$
\bea
 \Gamma_{B_{2a}} &=& - \f{4}{k^2} \int d^7 z\,\, \bar Q Q\,
\Big(\f{1}{\ep} - \gamma - 2\ln \f{\bar Q Q}{k \mu}\Big)
+ \f{8\ln2}{k^2} \int d^7 z\,\bar Q Q \nn \\
&&+\f{4}{k^2 \a^2} \int d^7 z\,
   \bar Q(z)\Big(\f{1}{\ep}-\gamma\Big) Q(z)\,.
\eea Note that it contains divergent quantum contributions which
appear as the pole $\frac1\varepsilon$. These divergent
contributions should be removed by adding the corresponding
counterterm to the classical action after computing all divergent
two-loop diagrams.

\subsubsection{Diagram $\Gamma_{B_{2b}}$}
The diagram $\Gamma_{2b}$ in Fig.\ \ref{Fig2} corresponds to the
following analytic expression \bea
 \Gamma_{B_{2b}}&=& 64\int d^7 z_1\,d^7 z_2\,d^7 z_3\,
    \bar Q Q\, G_{+-}(z_2,z_3) G_{+-}(z_1,z_2)  G(z_1,z_2)
    G(z_1,z_3)\,.
\eea Integrating by parts the covariant spinor derivatives which are
present in the (anti)chiral Green function $G_{+-}(z_1,z_2)$ we
collect them on the product $G(z_1,z_2) G(z_1,z_3)$ and then apply
the following identity \bea
 D^2_1\bar D^2_1 G(z_1,z_2) G(z_1,z_3)|_{\t_2=\t_1}&=&
 \bar D^2 G(z_1,z_2) D^2 G(z_1,z_3) +  D^2 G(z_1,z_2) \bar D^2 G(z_1,z_3)  \\
 && - 2 D^\b \bar D^\a G(z_1,z_2) D_\b \bar D_\a G(z_1,z_3)
 + G(z_1,z_2)D^2\bar D^2 G(z_1,z_3)\,.\nn
\eea Here we omit all terms which are equal to zero after the
integration over $\t_2$. After that we integrate by parts the
covariant spinor derivatives contained in $ G_{+-}(z_2,z_3)$ and get
\bea
 \Gamma_{B_{2b}}&=& 4\int d^7 z_1\,d^7 z_2\,d^7 z_3\,
    \bar Q Q\,\f{1}{\square}\d^7(z_3-z_2) \f{1}{\square}\d^7(z_1-z_2) \nn \\
&&   \times\f{\bar D^2_3 D^2_3}{16}  \bigg( \bar D^2 G(z_1,z_2) D^2
G(z_1,z_3)
+ D^2 G(z_1,z_2) \bar D^2 G(z_1,z_3) \nn \\
&& - 2 D^\b \bar D^\a G(z_1,z_2) D_\b \bar D_\a G(z_1,z_3)
 +\, G(z_1,z_2)D^2\bar D^2 G(z_1,z_3)  \bigg) \nn \\
&=& -\f{(16 \pi)^2}{k^2 \a^2}\int d^7 z_1\,d^3 x_2\, \bar
Q(x_1,\t_1) Q(x_2,\t_1) \f{1}{\square}\d^3(x_1-x_2)
\f{1}{\square}\d^3(x_1-x_2) \f{1}{\square}\d^3(x_1-x_2)
\nn \\
&& -\f{ (16\pi)^2}{k^2}\int d^7 z_1\,d^3 x_2\,d^3 x_3\, \bar Q Q
\f{1}{\square}\d^3(x_3-x_2) \f{1}{\square}\d^3(x_1-x_2) \nn \\
&&\qquad\times \f{1}{\square+\f{4\pi^2 M^2}{k^2}}\d^3(x_1-x_2)
\f{\square}{\square+\f{4\pi^2 M^2}{k^2}}\d^3(x_1-x_3) \nn\\
&&+\f{2(4\pi)^4}{k^4}\int d^7 z_1\,d^3 x_2\,d^3 x_3\, \bar Q Q\, M^2
\f{1}{\square}\d^3(x_3-x_2) \f{1}{\square}\d^3(x_1-x_2) \nn \\
&&\qquad\times \f{\pr^m}{\square(\square+\f{4\pi^2
M^2}{k^2})}\d^3(x_1-x_2) \f{\pr_m}{\square+\f{4\pi^2
M^2}{k^2}}\d^3(x_1-x_3) \label{B14_}
 \eea
Here we have done the integrals over Grassmann variables  $\t_3$ and
$\t_2$ and used the properties of the gauge superfield propagator
\eq{prG1}, \eq{prG2} and  \eq{G1=0}--\eq{G5=0}. Next, we pass to the
momentum representation in (\ref{B14_})
 \bea
 \Gamma_{B_{2b}}&=& \f{256\pi^2}{k^2 \a^2} \int d^4 \t \int
\f{d^3l}{(2\pi)^3}\bar Q(l,\t) Q(-l,\t)
\, \int\f{d^3p d^3q}{(2\pi)^6}\f{1}{(p-l)^2 q^2 (p+q)^2} \nn\\
&+& \f{256\pi^2}{k^2}\int d^7 z\,\, \bar Q Q\,
   \int \f{d^3 p\, d^3 q}{(2\pi)^6 }\,
       \f{1}{(p+q)^2(p^2-\f{4\pi^2 M^2}{k^2})(q^2-\f{4\pi^2 M^2}{k^2})} \nn\\
&+&  \f{256\pi^4}{k^4}\int d^7 z\, \bar Q Q\, M^2
   \int \f{d^3 p\, d^3 q}{(2\pi)^6 }
       \f{2p\cdot q}{p^2(p^2-\f{4\pi^2 M^2}{k^2})(p+q)^2 q^2(q^2-\f{4\pi^2 M^2}{k^2})}
 \eea
and compute the momentum integrals using \eq{I2}, \eq{I3} and
\eq{I4} \bea
 \Gamma_{B_{2b}}&=&
-\f{8}{k^2 \a^2} \int d^7 z\,
   \bar Q(z)\bigg(\f{1}{\ep}-\gamma\bigg) Q(z) \nn\\
&& - \f{8}{k^2} \int d^7 z\, \bar Q Q\, \bigg(\f{1}{\ep} - \gamma -
2\ln \f{\bar Q Q}{k \mu}\bigg)
 - \f{4(1-2\ln2)}{k^2} \int  d^7 z\,\bar Q Q\,.
\eea In this expression we omitted the terms containing the operator
$\ln\square$ since they do not contribute to the effective K\"ahler
potential.

\subsubsection{Diagram $\Gamma_{B_{2c}}$}
The sum of two diagrams $\Gamma_{2c}$ in Fig.\ \ref{Fig2}
corresponds to the following analytic expression \bea
 \Gamma_{B_{2c}}&=& - 32\int d^7 z_1\,d^7 z_2\,d^7 z_3\,d^7 z_4\,
    \bar Q Q\, G_{-+}(z_1,z_2)G_{+-}(z_3,z_4) G_{+-}(z_1,z_3)\nn \\
&& \times\left[ G(z_1,z_3) G(z_2,z_4) + G(z_1,z_4) G(z_2,z_3)
\right]\,. \la{A.15} \eea Let us consider the contributions from the
first term in the brackets in \eq{A.15}. We integrate by parts the
covariant spinor derivatives contained in $G_{+-}(z_2,z_1)$ and
$G_{-+}(z_4,z_3)$ and apply the identity \eq{prG2}
 \bea
 &&- 32\int d^7 z_1\,d^7 z_2\,d^7 z_3\,d^7 z_4\,
    \bar Q Q\,\nn \\ &&
    \qquad \times G_{+-}(z_1,z_3) G(z_1,z_3)%
     \f{1}{\square}\d^7(z_2-z_1)\f{1}{\square}\d^7(z_4-z_3)
    \f{\bar D^2_4  D^2_4}{16}\f{D^2_2 \bar D^2_2}{16}G(z_2,z_4) \nn \\
  && = \f{64i\pi}{k\a}\int d^7 z_1\,d^7 z_2\,d^7 z_3\,d^7 z_4\,
    \bar Q Q\, \nn  \\ &&
   \qquad \times G_{+-}(z_1,z_3) G(z_1,z_3)%
 \f{1}{\square}\d^7(z_2-z_1)\f{1}{\square}\d^7(z_4-z_3)
    \f{\bar D^2_4  D^2_4}{16}\d_-(z_2,z_4) = 0\,.
 \eea
This expression vanishes since there is the chiral delta-function
$\d_-(z_2,z_4)$ integrated over the full superspace.

The contribution from the second term in the brackets in \eq{A.15}
can be calculated in a similar way
 \bea
 \Gamma_{B_{2c}}&=& - 32\int d^7 z_1\,d^7 z_2\,d^7 z_3\,d^7 z_4\,
    \bar Q Q\, G_{-+}(z_1,z_2)G_{+-}(z_3,z_4)
    G_{+-}(z_1,z_3)G(z_1,z_4) G(z_2,z_3) \nn \\
&=& - 32\int d^7 z_1\,d^7 z_2\,d^7 z_3\,d^7 z_4\,
    \bar Q Q\,\f{1}{\square}\d^7(z_2-z_1) \f{1}{\square}\d^7(z_4-z_3)
    \,G_{+-}(z_1,z_3)     \nn\\
&& \qquad \times \f{ D^2_2 \bar D^2_2}{16} G(z_2,z_3)
  \f{\bar D^2_4  D^2_4}{16} G(z_1,z_4)\,.
 \eea
Here we integrated by parts the covariant spinor derivatives
contained in $ G_{-+}(z_1,z_2)$ and $ G_{+-}(z_3,z_4)$. Then we do
the integration over $\t_2$ and apply the identity \eq{prG2}
 \bea
 \Gamma_{B_{2c}}&=&
- \f{16i\pi}{k\a}\int d^7 z_1\,d^3 x_2\,d^7 z_3\,d^7 z_4\,
    \bar Q Q\,\f{1}{\square}\d^3(x_2-x_1) \f{1}{\square}\d^7(z_4-z_3)
 \f{1}{\square}\d^7(z_1-z_3) \nn\\
&& \qquad \times \f{ D^2_1 \bar D^2_1}{16}\bigg(
D^2(x_2,\t_1)\d^4(\t_1-\t_3) \d^3(x_2-x_3)\,\,
  \f{\bar D^2_4  D^2_4}{16} G(z_1,z_4)\bigg)\,.
 \eea
The non-zero contribution arises only when the operator $D^2_1$ acts
on $G(z_1,z_4)$. After integration over $\t_4$ and $\t_3$ we obtain
two bosonic delta-functions $\d^3(x_2-x_3)$ and $\d^3(x_1-x_4)$
which allow us to do the integration over $x_3$ and $x_4$ as well
\bea
 \Gamma_{B_{2c}}&=&
 \f{128\pi^2}{k^2\a^2}\int d^7 z_1\,d^3 x_2\,
    \bar Q Q\,\f{1}{\square}\d^3(x_2-x_1) \f{1}{\square}\d^3(x_1-x_2)
 \f{1}{\square}\d^3(x_1-x_2) \nn \\
 &=&\f{4}{k^2 \a^2} \int d^7 z\,
   \bar Q(z)\left(\f{1}{\ep}-\gamma \right) Q(z)\,.
 \eea
Here we applied the formula (\ref{I2}) to compute the corresponding
momentum integral.

\section{Momentum integrals}
In this Appendix we give the list of momentum integrals which appear
in one- and two-loop Feynman graphs considered in the present paper.
Some of these integrals can be found in the textbook
\cite{Frampton}: \bea
 J(m) &=& \int \f{d^3 p}{(2\pi)^3} \f{1}{(p^2) (p^2 - m^2)} =
 -\f{i}{4 \pi\,|m|}\,, \la{J} \\
 I(p,m)&=& \int \f{d^3 q}{(2\pi)^3} \f{1}{(p+q)^2 (q^2-m^2)} =
 \f{1}{8\pi}\int_0^1\f{ dx}{\sqrt{1-x}\,(p^2 x - m^2)^{1/2}}  \la{I}\\
 I_1&=&\int \f{d^3 p}{(2\pi)^3} \f{I(p,m)}{p^2-m^2} =
 - \f1{32\pi^2}\f{\Gamma(\ep)}{m^{2\ep}} +
\f{\ln2}{16 \pi^2}\,, \la{I2} \\
I_2&=&\int\f{d^3 p }{(2\pi)^3}\f{I(p,0)}{(p-l)^2} =
-\f{1}{32\pi^2}\f{\Gamma(\ep)}{(-l^2)^{\ep}} \la{I3}\,, \\
I_3&=&\int \f{d^3 p\, d^3 q}{(2\pi)^6 }
       \f{2p\cdot q}{p^2(p^2-m^2)(p+q)^2 q^2(q^2-m^2)}
=- \f{1+2\ln2}{16\pi^2\,m^2} \la{I4}\,.
 \eea
Here $\ep$ is the parameter of dimensional regularization,
$d=3-2\varepsilon$, $\ep \to 0$. The divergent parts of these
integrals can be singled out by the standard series expansion of the
$\Gamma$-function \be \frac{\Gamma(\epsilon)}{m^{2\epsilon}}=
\frac1\epsilon-\gamma-\ln m^2+O(\epsilon)\,, \ee where
$\gamma=0,577...$ is the Euler constant.


\begin{thebibliography}{99}
%
\bibitem{book}
  I.~L.~Buchbinder, S.~M.~Kuzenko,
  {\it Ideas and methods of supersymmetry and supergravity: Or a walk through
  superspace}, Bristol, UK: IOP (1998) 656 p.
%
\bibitem{BKY}
  I.~L.~Buchbinder, S.~Kuzenko and Z.~Yarevskaya,
  {\it Supersymmetric effective potential: Superfield approach},
  Nucl.\ Phys.\ B\ {\bf 411} (1994) 665.
%
\bibitem{BKPY}
  I.~L.~Buchbinder, S.~M.~Kuzenko, A.~Y.~Petrov and Z.~V.~Yarevskaya,
  {\it Superfield effective potential}, {\tt hep-th/9501047}.
%
\bibitem{BKP}
  I.~L.~Buchbinder, S.~M.~Kuzenko and A.~Y.~Petrov,
  {\it Superfield effective potential in the two loop approximation},
  Phys.\ Atom.\ Nucl.\  {\bf 59} (1996) 148,
   [Yad.\ Fiz.\  {\bf 59} (1996) 157].
%
\bibitem{BCP1}
  I.~L.~Buchbinder, M.~Cvetic and A.~Y.~Petrov,
  {\it Implications of decoupling effects for one loop corrected effective actions from superstring theory},
  Mod.\ Phys.\ Lett.\ A\ {\bf 15} (2000) 783,
  {\tt  hep-th/9903243}.
%
\bibitem{BCP2}
  I.~L.~Buchbinder, M.~Cvetic and A.~Y.~Petrov,
  {\it One loop effective potential of N=1 supersymmetric theory and decoupling effects},
  Nucl.\ Phys.\ B\ {\bf 571} (2000) 358,
  {\tt hep-th/9906141}.
%
\bibitem{Brignole} A. Brignole, {\it One-loop K\"ahler potential in non-renormalizable
theories}, Nucl. Phys. B {\bf 579} (2000) 101, {\tt hep-th/0001121}.
%
\bibitem{deWit1996}
  B.~de Wit, M.~T.~Grisaru and M.~Ro\v cek,
  {\it Nonholomorphic corrections to the one loop N=2 superYang-Mills action},
  Phys.\ Lett.\ B {\bf 374} (1996) 297,
  {\tt hep-th/9601115}.
%
\bibitem{PW}
  A.~Pickering and P.~C.~West,
  {\it The one-loop effective superpotential and nonholomorphicity},
  Phys.\ Lett.\ B {\bf 383} (1996) 54,
  {\tt hep-th/9604147}.
%
\bibitem{Grisaru96}
  M.~T.~Grisaru, M.~Ro\v cek and R.~von Unge,
  {\it Effective K\"ahler potentials},
  Phys.\ Lett.\ B {\bf 383} (1996) 415,
  {\tt hep-th/9605149}.
%
\bibitem{Nibbelink}
  S.~Groot Nibbelink and T.~S.~Nyawelo,
  {\it Two Loop effective Kahler potential of (non-)re\-nor\-ma\-li\-za\-ble supersymmetric models},
  JHEP {\bf 0601} (2006) 034,
  {\tt hep-th/0511004}.
%
\bibitem{Tyler} S. J. Tyler, {\it Studies of low-energy effective actions in
supersymmetric field theories}, PhD thesis, {\tt arXiv:1401.4814
[hep-th].}
%
\bibitem{Petrov}
  M. Gomes, A. C. Lehum, J. R. Nascimento, A. Yu. Petrov, A. J. da
  Silva,
  {\it The effective superpotential in the supersymmetric
  Chern-Simons theory with matter},
      Phys.\ Rev.\ D {\bf 87}  (2013) 027701, {\tt arXiv:1210.6863 [hep-th].}
%
\bibitem{Lehum}
  A.~F.~Ferrari, M.~Gomes, A.~C.~Lehum, J.~R.~Nascimento, A.~Y.~Petrov, E.~O.~Silva and A.~J.~da Silva,
  {\it On the superfield effective potential in three dimensions},
  Phys.\ Lett.\ B\ {\bf 678}  (2009) 500,
  {\tt arXiv:0901.0679 [hep-th]}.
%
\bibitem{BMS3} I.L. Buchbinder, B.S. Merzlikin,
 I.B. Samsonov, {\it Two-loop effective potentials in general N=2, d=3
chiral superfield model}, Nucl. Phys. B {\bf 860}  (2012) 87, {\tt
arXiv:1201.5579 [hep-th]}.

\bibitem{deBoer97}
  J.~de Boer, K.~Hori, Y.~Oz and Z.~Yin,
  {\it Branes and mirror symmetry in N=2 supersymmetric gauge theories in three-di\-men\-si\-ons},
  Nucl.\ Phys.\ B {\bf 502} (1997) 107,
  {\tt hep-th/9702154}.
%
\bibitem{deBoer}
  J.~de Boer, K.~Hori and Y.~Oz,
  {\it Dynamics of N=2 supersymmetric gauge theories in
  three-dimensions},  Nucl.\ Phys.\ B {\bf 500} (1997) 163,
  {\tt hep-th/9703100}.
%
\bibitem{AHISS}
  O.~Aharony, A.~Hanany, K.~A.~Intriligator, N.~Seiberg and M.~J.~Strassler,
  {\it Aspects of N=2 supersymmetric gauge theories in three-dimensions},
  Nucl.\ Phys.\ B {\bf 499} (1997) 67,
  {\tt hep-th/9703110}.
%
\bibitem{IS13}
  K.~Intriligator and N.~Seiberg,
  {\it Aspects of 3d N=2 Chern-Simons-matter theories},
  JHEP {\bf 1307} (2013) 079,
  {\tt arXiv:1305.1633 [hep-th]}.
%
\bibitem{1} I.~L.~Buchbinder, B.~S.~Merzlikin, I.~B.~Samsonov,
 {\it Two-loop low-energy effective action in Abelian supersymmetric
 Chern-Simons matter models}, Nucl.\ Phys.\ B {\bf 881} (2014) 42,
 {\tt arXiv:1311.5001 [hep-th]}.
%
\bibitem{2} I.~L.~Buchbinder, B.~S.~Merzlikin, I.~B.~Samsonov,
 {\it Two-loop low-energy effective actions in N=2 and N=4 three-dimensional
 SQED}, JHEP {\bf 07} (2013) 012, {\tt arXiv:1305.4815 [hep-th]}.
%
\bibitem{3} I.~L.~Buchbinder, N.~G.~Pletnev, I.~B.~Samsonov,
 {\it Background field formalism and construction of effective action
 for N=2, d=3 supersymmetric gauge theories}, Phys.\ Part.\ Nuclei {\bf 44}
 (2013) 234, {\tt arXiv:1206.5711 [hep-th]}.
%
\bibitem{4} I.~L.~Buchbinder, N.~G.~Pletnev, I.~B.~Samsonov,
 {\it Low-energy effective actions in three-dimensional extended
 SYM theories}, JHEP {\bf 01} (2011) 121, {\tt arXiv:1010.4967 [hep-th]}.
%
\bibitem{5} I.~L.~Buchbinder, N.~G.~Pletnev, I.~B.~Samsonov,
 {\it Effective action of three-dimensional extended supersymmetric
 matter on gauge superfield background}, JHEP {\bf 04} (2010) 124, {\tt arXiv:1003.4806 [hep-th]}.


\bibitem{Avdeev1}
  L.~V.~Avdeev, G.~V.~Grigorev and D.~I.~Kazakov,
  {\it Renormalizations in Abelian Chern-Simons field theories with matter},
  Nucl.\ Phys.\ B {\bf 382} (1992) 561.
%
\bibitem{Avdeev2}
  L.~V.~Avdeev, G.~V.~Grigorev and D.~I.~Kazakov,
  {\it Renormalizations in Abelian Chern-Simons field theories with matter},
  Nucl.\ Phys.\ B  {\bf 382} (1992) 561.
%
\bibitem{BILPSZ}
  I.~L.~Buchbinder, E.~A.~Ivanov, O.~Lechtenfeld, N.~G.~Pletnev, I.~B.~Samsonov and B.~M.~Zupnik,
  {\it Quantum N=3, d=3 Chern-Simons matter theories in harmonic superspace},
  JHEP {\bf 0910} (2009) 075,
  {\tt arXiv:0909.2970 [hep-th]}.
%
\bibitem{Seib88} N.~Seiberg, {\it Supersymmetry and
nonperturbative beta functions}, Phys.\ Lett.\ B {\bf206} (1988) 75.
%
\bibitem{anomaly1}
  A.~J.~Niemi and G.~W.~Semenoff,
  {\it Axial anomaly induced fermion fractionization and effective gauge theory
  actions in odd dimensional space-simes},
  Phys.\ Rev.\ Lett.\  {\bf 51} (1983) 2077.
%
\bibitem{anomaly2} A.~N.~Redlich,
{\it Gauge noninvariance and parity violation of
three-di\-men\-si\-onal fer\-mi\-ons}, Phys. Rev. Lett. {\bf 52}
(1984) 18.
%
\bibitem{anomaly3} A.~N.~Redlich,
{\it Parity violation and gauge noninvariance of the effective gauge
field action in three dimensions}, Phys. Rev. D {\bf 29} (1984)
2366.
%
\bibitem{DeWitt} B.~S.~DeWitt, {\it Dynamical Theory of Groups and
Fields}, Gordon and Breach, 1965.
%
\bibitem{GGRS} S.~J.~Gates, M.~T.~Grisaru, M.~Ro\v cek, W.~Siegel,
{\it Superspace or One Thousand and One Lessons in Supersymmetry},
Benjamin/Cummings, Reading, MA, 1983, 548 p.
%
\bibitem{NG} S.~J.~Gates Jr. and H.~Nishino, {\it Remarks on N=2
supersymmetric Chern-Simons theories}, Phys. Lett. B {\bf 281}
(1992) 72.
%
\bibitem{Kuz07} S.~M.~Kuzenko and S.~J.~Tyler, {\it Supersymmetric Euler-Heisenberg
effective action: Two-loop results}, JHEP  {\bf 0705} (2007) 081,
{\tt hep-th/0703269}.
%
\bibitem{OV} A.A. Ostrovsky, G.A. Vilkovisky, {\it The covariant
effective action in QED. One-loop magnetic moment}, J. Math. Phys.
{\bf 29} (1988) 702.
%
\bibitem{Argyres}
  P.~C.~Argyres, M.~R.~Plesser and N.~Seiberg,
  {\it The moduli space of vacua of N=2 SUSY QCD and duality in N=1 SUSY QCD},
  Nucl.\ Phys.\ B {\bf 471} (1996) 159,
  {\tt hep-th/9603042}.
%
\bibitem{IntSeib}
  K.~A.~Intriligator and N.~Seiberg,
  {\it Mirror symmetry in three-dimensional gauge theories},
  Phys.\ Lett.\ B {\bf 387} (1996) 513,
  {\tt hep-th/9607207}.
%
\bibitem{Frampton}
  P.~H.~Frampton,
  {\it Gauge Field Theories: Third Revised and Improved Edition},
  Weinheim, Germany: Wiley-VCH (2008) 353 p.


\end{thebibliography}
\end{document}